\documentstyle[twoside,fleqn,espcrc2,psfig]{article}
\def\ib#1#2#3{           {\it ibid. }{\bf #1} (19#2) #3}

\def\nps#1#2#3{          {\it Nucl. Phys. B (Proc.Suppl.) }
                         {\bf #1} (19#2) #3} 

\def\pl#1#2#3{           {\it Phys. Lett. }{\bf #1} (19#2) #3}
\def\pr#1#2#3{        {\it Phys.Rev.}{\bf #1} (19#2) #3}

\def\n.c.#1#2#3{         {\it Nuovo Cim. }{\bf #1} (19#2) #3}
\def\r.n.c.#1#2#3{       {\it Riv. del Nuovo Cim. }{\bf #1} (19#2) #3}

\def\jetp#1#2#3{         {\it Sov. Phys. JETP }{\bf #1} (19#2) #3}
\def\jetpl#1#2#3{         {\it JETP Lett. }{\bf #1} (19#2) #3}

\def\ppnp#1#2#3{           {\it Prog. Part. Nucl. Phys. }{\bf #1} (19#2) #3}

\def\eq#1{{eq. (\ref{#1})}}
\def\ne{\hbox{$\nu_e$ }}

\newcommand{\AmS}{{\protect\the\textfont2
  A\kern-.1667em\lower.5ex\hbox{M}\kern-.125emS}}

\title{Low-energy Anti-neutrinos from the Sun}
\author{\underline{V.B. Semikoz}\thanks{On leave from  
IZMIRAN, Troitsk, Moscow region, 142092, Russia},
S. Pastor\thanks{E-mail: sergio@flamenco.ific.uv.es}
and J.W.F. Valle\thanks{E-mail: valle@flamenco.ific.uv.es~
Work supported by DGICYT under
grants PB95-1077 and SAB95-506 (V.B.S.), by the TMR network grant
ERBFMRXCT960090 and by INTAS grant 96-0659 of the European
Union. S.P. was supported by Conselleria d'Educaci\'o i Ci\`encia of
Generalitat Valenciana. V.B.S. also acknowledges the support of RFFR
through grants 97-02-16501 and 95-02-03724.}
\address{Instituto de F\'{\i}sica Corpuscular - C.S.I.C., 
Departament de F\'{\i}sica Te\`orica, Universitat de Val\`encia\\
46100 Burjassot, Val\`encia, SPAIN}}

\begin{document}

\begin{abstract}
If neutrino conversions within the Sun result in partial
polarization of initial solar neutrino fluxes, then a new opportunity
arises to observe the $\bar{\nu}_e$'s in future neutrino experiments
in the low energy region (such as BOREXINO or HELLAZ) and thus to
probe the Majorana nature of the neutrinos. The
$\nu_e \rightarrow \bar{\nu}_e$ conversions may take place for low
energy solar neutrinos while being
unobservable at the Kamiokande and Super-Kamiokande experiments.
\end{abstract}

\maketitle

Finding a signature for the Majorana nature of neutrinos or,
equivalently, for the violation of lepton number in Nature is a
fundamental challenge in particle physics \cite{fae}. All attempts for
distinguishing Dirac from Majorana neutrinos {\em directly} in
laboratory experiments have proven to be a hopeless task, due to the
V-A character of the weak interaction, which implies that all such
effects vanish as the neutrino mass goes to zero. We suggest an
alternative way in which one might probe for the possibility of
$L$-violation which is not {\em directly} induced by the presence
of a Majorana mass. Although, Majorana masses will be
required at some level, but the quantity which is directly involved is the
transition amplitude for a \ne to convert into an $\bar{\nu}_e$ inside the
Sun. 

Here we propose to probe for the possible existence of
$L$-violating processes in the solar interior that can produce an
$\bar{\nu}_e$ component in the neutrino flux.  The idea is that, even
though the nuclear reactions that occur in a normal star like our Sun
do not produce directly right-handed active neutrinos ($\bar{\nu}_a$)
these may be produced by combining the chirality-flipping transition
$\nu_{eL} \rightarrow \bar{\nu}_{a R}$ with the standard
chirality-preserving MSW conversions $\nu_{eL} \rightarrow \nu_{\mu
L}$ through cascade conversions like $\nu_{eL} \rightarrow
\bar{\nu}_{\mu R} \rightarrow \bar{\nu}_{eR}$ or $\nu_{eL} \rightarrow
\nu_{\mu L} \rightarrow \bar{\nu}_{eR}$.  These conversions arise as a
result of the interplay of two types of mixing \cite{akhpetsmi}: one
of them, matter-induced flavour mixing, leads to MSW resonant
conversions which preserve the lepton number $L$, whereas the other is
generated by the resonant interaction of a Majorana neutrino
transition magnetic moment with the solar magnetic field
\cite{LAM}. This violates the $L$ symmetry by two units ($\Delta L =
\pm 2$) and is an explicit signature of the Majorana nature of the
neutrino \cite{BFD}.

We consider neutrino-electron scattering in future underground solar
neutrino experiments in the low-energy region, below the threshold for
$\bar{\nu}_e + p \rightarrow n +e^+$, such as is the case for $pp$ or
$^7$Be neutrinos. These should be measured in future
real-time experiments such as HELLAZ \cite{hellaz} or BOREXINO
\cite{borexino} which will have low energy thresholds ($100$ keV and
$250$ keV, respectively). BOREXINO is designed to take
advantage of the characteristic shape of the electron recoil energy
spectrum from the $^7$Be neutrino line, while 
the HELLAZ experiment is intended to measure the
fundamental neutrinos of the $pp$ chain.

The complete expression for the differential cross section of the weak
process $\nu e \rightarrow \nu e$, as a function of the electron
recoil energy $T$, in the massless neutrino limit, can be written as
%
%
%
\begin{equation}
\label{totdsigma}
\frac{d \sigma}{dT} (\omega,T)= \frac{2 G_F^2 m_e}{\pi} \Bigl [
P_e h(g_{eL},g_R) 
\end{equation}
$$
+ P_{\bar{e}} h(g_R,g_{eL})
+ P_a h(g_{aL},g_R)
+ P_{\bar{a}} h(g_R,g_{aL})\Bigr ]
$$
where
$h(a,b) \equiv  a^2 + b^2 (1- T/\omega)^2 -
ab m_e T/\omega^2$
and $g_{eL} = \sin^2 \theta_W+0.5$ , $g_{a L} = \sin^2 \theta_W-0.5$
($a=\mu,\tau$) and $g_R = \sin^2 \theta_W$ are the weak couplings of
the Standard Model, and $\omega$ is the energy of the incoming
neutrino. The parameter $P_e$ in the equation above is the survival
probability of the initial left-handed electron neutrinos, while
$P_{\bar{e}}$, $P_a$ and $P_{\bar{a}}$ are the appearance
probabilities of the other species, that may arise in the Sun as a
result of the processes $\nu_{eL} \to \bar{\nu}_{eR}$, $\nu_{eL}\to
\nu_{aL}$ or $\nu_{eL} \to \bar{\nu}_{aR}$, respectively. These
parameters obey the unitarity condition $P_e(\omega) +
P_{\bar{e}}(\omega) + P_a(\omega) + P_{\bar{a}}(\omega) = 1$.  In
general they are obtained from the complete $4\times 4$ evolution
Hamiltonian describing the evolution of the neutrino system
\cite{BFD}. They depend on the neutrino energy $\omega$,
on the solar magnetic field through the parameter $\mu_\nu B_\perp$
and on the neutrino mixing parameters $\Delta m^2$, $\sin ^22\theta$.

In the $L$-violating processes one has in general all four
contributions shown in \eq{totdsigma}. In contrast, in the case where
lepton number is conserved (like in MSW conversions), the solar
neutrino flux will consist of {\em neutrinos}, so only the first and
third terms in \eq{totdsigma} contribute.  It follows that the
differential cross section will be {\em different} in the case where
$\nu_e$'s from the Sun get converted to electron $\bar{\nu}_e$'s. Is
it possible to measure this difference in future neutrino experiments?

The relevant quantity to be measured in neutrino scattering
experiments is the energy spectrum of events, namely
\begin{equation}
\label{spectrum}
\frac{dN_\nu}{dT} = N_e \sum_{i} \phi_{0i} 
\int^{\omega_{max}}_{\omega_{min}(T)}
d \omega \lambda_i(\omega) 
\langle \frac{d \sigma}{dT} (\omega,T) \rangle
\end{equation}
where $d \sigma/dT$ is given in \eq{totdsigma} and $N_e$ is the number
of electrons in the fiducial volume of the detector
\footnote{For simplicity we have taken in \eq{spectrum} 
the efficiency of the detector as unity for energies above the
threshold.}.  The sum in the above equation is done over the solar
neutrino spectrum, where $i$ corresponds to the different reactions
$i= pp$, $^7$Be, $pep$, $^8$B $\ldots$, characterized by an integral
flux $\phi_{0i}$ and a differential spectrum $\lambda_i(\omega)$ (for
neutrinos coming from two-body reactions, one has $\lambda_i(\omega) =
\delta(\omega - \omega_i)$). The lower limit for the neutrino energy
is $\omega_{min} (T) = 1/2(T + \sqrt{T^2 +2m_eT})$, while the upper
limit $\omega_{max}$ corresponds to the maximum neutrino energy. In
order to take into account the finite resolution in the measured
electron recoil energy, we perform a Gaussian average of the cross
section, indicated by $\langle \ldots \rangle$ in \eq{spectrum}.  For
further details, see ref. \cite{paper}.

%

\begin{figure*}
\centerline{\protect\hbox{\psfig{file=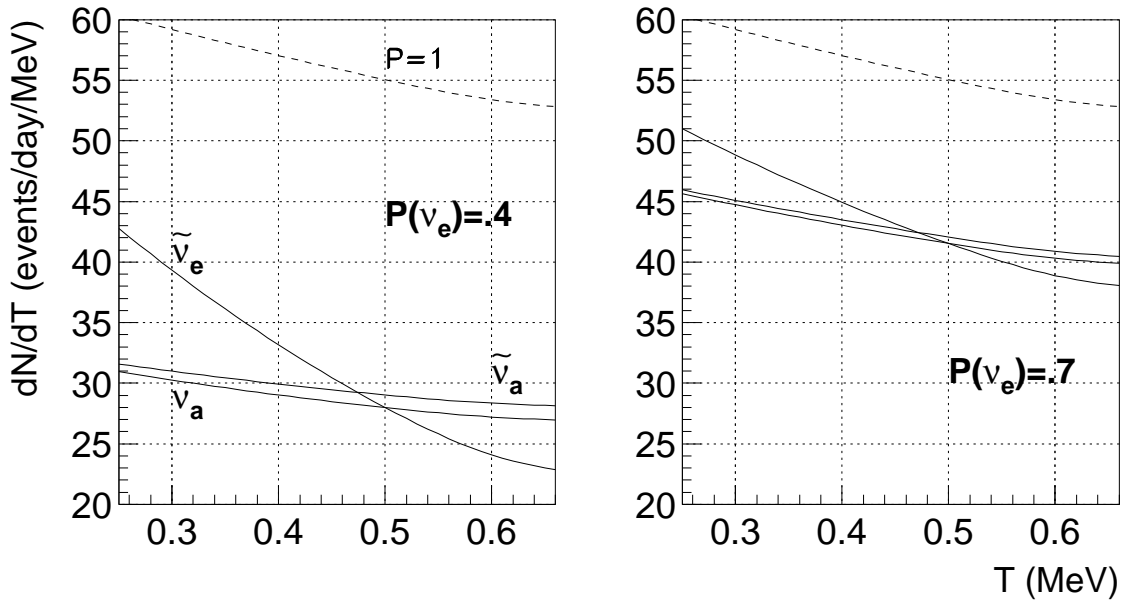,height=6.25cm}}}
\caption{Energy spectrum of events corresponding to $^7$Be solar
neutrinos for the BOREXINO experiment. Different cases are shown,
as described in the text.}
\label{fig1}
\end{figure*}

We have calculated the averaged energy spectrum of events
for the two experiments in the simple case where the
parameters $P_i$ do not depend on the neutrino energy.  Here we
present in figure
\ref{fig1} our results for BOREXINO.
The upper line corresponds to
the case where one has no neutrino conversions ($P_e=1$). When
electron anti-neutrinos are present in the solar flux the results are
the lines labelled with $\bar{\nu}_e$, calculated for the indicated
value of $P_e$ and the corresponding amount of $\bar{\nu}_a$. The
cases of $\nu_e \rightarrow \nu_{\mu,\tau}$ and $\nu_e \rightarrow
\bar{\nu}_{\mu,\tau}$ are the lower lines with labels $\nu_a$ and
$\bar{\nu}_a$, respectively.

One can see from figure \ref{fig1} that it is possible to distinguish
the case with $\bar{\nu}_e$ considering the behaviour of the cross
section for low energies. It is the {\em slope} of the measured
spectrum the key for recognizing the presence of $\bar{\nu}_e$'s
in the solar neutrino flux, and correspondingly the
presence of $L$-violating processes which can only exist if neutrinos
are Majorana particles.  

The shortcoming of the above discussion is that 
we have neglected the energy dependence of the
physical parameters $P_i$. One must calculate
the averaged $\nu-e$ cross section using analytical expressions for
$P_i=P_i(\omega)$.  However, since the $^7$Be
neutrinos are mono-energetic, whatever the mechanism that produces the
deficit is, their survival probability will take on a constant value
$P_e(\omega_{Be})$. Therefore one can apply directly the results we
have obtained for constant $P_i$ for the range of electron recoil
energy where the contribution of $^7$Be neutrinos dominates. In
ref. \cite{paper} we give a discussion of the experimental uncertainties.

There are however stringent bounds on the presence of solar 
$\bar{\nu}_e$'s in the high energy region ($^8$B). These would interact
within the detector through the process $\bar{\nu}_e + p \rightarrow n
+e^+$.  This process, which has an energy threshold of $E_\nu = m_n
-m_p +m_e \simeq 1.8$ MeV, has not been found to occur in the
Kamiokande experiment \cite{barbieri91,raghavan91}, nor in the very
recent data from Super-Kamiokande \cite{fiorentini97}.  Also the
results from the liquid scintillation detector (LSD) are negative
\cite{aglietta96}.  

As we show in ref. \cite{paper} for the specific scenario presented in
\cite{akhpetsmi}, the co-existence of a suppressed production of
high-energy $\bar{\nu}_e$'s and a sizeable flux of anti-neutrinos at
energies below $1.8$ MeV can be easily understood theoretically. The
resonance in the the $\nu_{eL} \rightarrow \bar{\nu}_{eR}$ conversions
can lie in the energy region below $1$ MeV (relevant for HELLAZ or
BOREXINO) provided that the neutrino parameters have reasonable
values, so that the conversion probability is small for energies
$\omega \gg 1$ MeV. Therefore the anti-neutrino flux would be {\em
hidden} in the background and therefore unobservable in
Super-Kamiokande.
%
%


Our conclusion is that neutrino conversions within the Sun can result
in partial polarization of the initial fluxes, in such a way as to
produce a sizeable $\bar{\nu}_e$ component without conflicting present
Super-Kamiokande data. The observation of $\bar{\nu}_e$'s from the Sun
in future neutrino experiments in the low energy region could lead to
the conclusion that the neutrinos are Majorana particles.



\end{document}